# Label-Efficient Self-Training for Attribute Extraction from Semi-Structured Web Documents

Ritesh Sarkhel*
The Ohio State University
Columbus, Ohio, USA

Binxuan Huang
Amazon Science
Seattle, Washington, USA

Colin Lockard
Amazon Science
Seattle, Washington, USA

Prashant Shiralkar
Amazon Science
Seattle, Washington, USA

## ABSTRACT
Extracting structured information from HTML documents is a long-studied problem with a broad range of applications, including knowledge base construction, faceted search, and personalized recommendation. Prior works rely on a few human-labeled web pages from each target website or thousands of human-labeled web pages from some seed websites to train a transferable extraction model that generalizes on unseen target websites. Noisy content, low site-level consistency, and lack of inter-annotator agreement make labeling web pages a time-consuming and expensive ordeal. We develop LEAST – a **La**bel-**E**fficient **S**elf-**T**raining method for Semi-Structured Web Documents to overcome these limitations. LEAST utilizes a few human-labeled pages to pseudo-annotate a large number of unlabeled web pages from the target vertical. It trains a transferable web-extraction model on both human-labeled and pseudo-labeled samples using self-training. To mitigate error propagation due to noisy training samples, LEAST re-weights each training sample based on its estimated label accuracy and incorporates it in training. To the best of our knowledge, this is the first work to propose end-to-end training for transferable web extraction models utilizing only a few human-labeled pages. Experiments on a large-scale public dataset show that using less than ten human-labeled pages from each seed website for training, a LEAST-trained model outperforms previous state-of-the-art by more than 26 average F1 points on unseen websites, reducing the number of human-labeled pages to achieve similar performance by more than 10x.

---

*This work was done during the first author's internship at Amazon

## 1 INTRODUCTION
Semi-structured websites are rich sources of high-quality information in many domains. Extracting structured information from these websites is a well-studied problem with a broad range of applications, including knowledge-base construction [4, 9], e-commerce product search [2], recommendation systems [31], knowledge-aware question answering [7], and more.

**Motivation.** We focus on a Information Extraction (IE) task from semi-structured *detail pages* in this paper. A detail page is a web page that corresponds to a single entity, typically the topic entity of the page e.g., the IMDB page of a movie or a product page on Amazon. Take the detail page in Fig. 1, for example. We want to extract a predefined set of attributes $A' = \{$*title, director, genre, mpaa-rating*$\}$ from this page. For each attribute in $A'$, we want to identify the corresponding text-span(s) appearing on that page. Compared to unstructured text, for which we can model this task as a sequence-tagging problem [21], semi-structured data format and rich page layout make this task more challenging for detail pages. To execute this task, a conventional extraction method such as *wrapper induction* [16] takes some human-annotated pages from each target website as input, derives XPath-based extraction rules or wrappers, and uses them to identify text-spans corresponding to each attribute. Although this approach yields high accuracy [12], it does not generalize well for unseen websites as it requires each target website to be labeled by human annotators. This makes wrapper induction methods prohibitively expensive for scenarios where we are interested in extracting from thousands of websites in a target vertical. Therefore, our interest lies in developing a label-efficient training method for web-extraction models that generalize well on unseen websites.

**State-of-the-art.** Contemporary researchers have explored several different avenues to reduce the number of human-labeled pages required to train a web-extraction model. Some recent works [11, 19] have explored distant supervision approaches where they align an external knowledge-base against each target website. These approaches, however, have some drawbacks. *First*, a comprehensive knowledge-base is not available for many emerging domains. *Second*, an extraction model learned this way does not generalize well on long-tail entities. In search of transferable web-extraction models, researchers [13, 20] have also utilized visual features of a DOM node (e.g., bounding-box coordinates) to identify website-invariant patterns. Encoding visual patterns, however, requires a computationally expensive rendering process and additional storage for resources such as images, CSS, and JavaScript files that can quickly get out-of-date. To address this, researchers [18, 36] have developed neural extraction models that jointly encode local and pairwise relationships between neighboring nodes from human-labeled pages of some seed websites. They have reported state-of-the-art extraction performance on unseen websites. These methods, however, require thousands of human-labeled pages from each seed website to train their models. Noisy multilingual content, low page-level consistency, and lack of inter-annotator agreement make labeling detail pages with sufficiently high accuracy a time-consuming process. If the number of human-labeled pages is small, the performance of these models worsens significantly. We address this gap by answering two key questions: (a) *how to train an extraction model that not only performs well on the seed websites but also on unseen websites from the same vertical?*, (b) *how to achieve (a) using only a few human-labeled pages for training?*

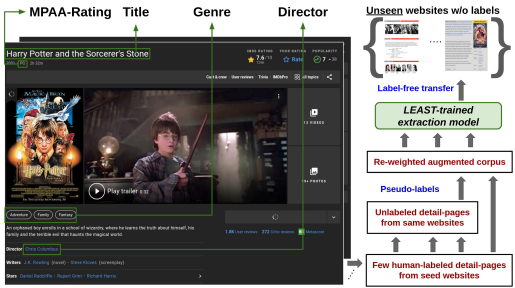

**Figure 1: LEAST is an end-to-end training framework for transferable web extraction using only a few human-labeled and a large number of unlabeled detail pages from a target vertical. In this example, a LEAST-trained model extracts text-spans corresponding to each attribute in $A'$={*title, director, genre, mpaa-rating*} from previously unseen websites from the 'movie' vertical.**

**Task and framework overview.** Given a set of attributes $A'$ and a website $w$ in vertical $v$, our goal is to learn a model $f(:, \theta)$ that extracts text-span(s) corresponding to each attribute $a \in A'$ from $w$ using only a few human-labeled pages from some seed websites to train $f(:, \theta)$. To this end, we propose a label-efficient self-training method that successfully addresses the two research questions mentioned above. Our main contributions are as follows.

*(i) Self-training with a few human-labeled pages:* We propose a self-training method for transferable web-extraction models. We co-train a teacher and a student model [28] on a few human-labeled and a large number of pseudo-labeled pages from the same vertical using iterative knowledge exchange. Once training terminates, the final model undertakes a node-level classification task to identify the node-texts corresponding to each attribute $a \in A'$.

*(ii) Pseudo-labeled corpus construction:* In a traditional self-training framework, the teacher model is trained on some labeled samples first. The trained teacher model is then used to pseudo-annotate task-specific unlabeled samples. We train the student model on an augmented corpus that contains both human-labeled and pseudo-labeled samples. This results in gradual drifts [25] in model quality due to noisy pseudo-labeled samples. To circumvent this, NLP researchers [23, 32] have used pre-trained language models to initialize the base teacher model, fine-tuned it on a few human-labeled samples, and then used the fine-tuned teacher model for pseudo-labeling. Developing a comprehensive, structure-aware language model for detail pages is an exciting direction of future work. We take a slightly different approach in this paper. Our method works in tandem with multiple existing models, including pre-trained language models. As a naive transfer of teacher knowledge, trained on a few human-labeled samples may result in noisy pseudo-labels, we develop a *semi-supervised generative model* as a *supplementary supervision source* for pseudo-label generation. We combine this generative model with the teacher network to construct the student training corpus. We assume that detail pages of a website are populated by passing each tuple of an underlying relational table through a website-specific rendering function $R(.)$. Our generative model learns to invert $R(.)$ to identify the relational table for each seed website and then uses it as a distant supervision source to pseudo-annotate unlabeled pages from the target vertical (Section 3.1).

*(iii) Node-level adaptive re-weighting:* Following prior works [18, 36], we formulate the extraction task as a DOM-node classification problem. To mitigate error propagation due to training on noisy pseudo-labeled samples, we develop an uncertainty-aware learning framework that adaptively re-weights each DOM-node in the training corpus based on its estimated label accuracy and incorporates it as an uncertainty measure during training. Contrary to prior works [17] that follow similar strategies for image/text classification, we do not make any assumption about the content of an unlabeled page for re-weighting. Figure 2 shows the end-to-end framework that implements all of the components described above. We refer to this framework as LEAST.

**Summary of results.** Through comprehensive experiments on four separate verticals of the Structured Web Data Extraction (SWDE) dataset [13], we show that a LEAST-trained model significantly outperforms strong baselines on all verticals. Our results show that a LEAST-trained model is not only *transferable*, i.e., it generalizes to unseen websites better, but also more *label-efficient*, i.e., it achieves better end-to-end performance using fewer human-labeled pages for training. We observe that training on less than ten human-labeled pages from each seed website and a large number of pseudo-labeled pages from the same vertical, a LEAST-trained model outperforms previous state-of-the-art [36] by 26.77% average F1 points on unseen websites, reducing the amount of human-labeled pages required for comparable performance by more than 10x. We introduce some related background concepts and formalize the extraction task studied in this paper next.

## 2 BACKGROUND & PROBLEM FORMULATION

**Attribute extraction from websites.** We assume that a website $w_i$ is a collection of detail pages $w_i = \bigcup_j p_j$ that share similar page layouts. A detail page $p_j$ describes a single topic entity and can be represented as a DOM-tree $t_j$. This means that each element of a detail page corresponds to an XPath and a DOM node. Following prior work [18, 36], we formulate the extraction task we study in this paper as a DOM node classification problem. We only classify DOM nodes that have textual content and assume that each DOM node corresponds to at most one attribute [13]. Extracting the attribute-set $A' = \{title, director, genre, mpaa-rating\}$ from the detail page shown in Fig. 1, therefore, boils down to classifying each DOM node as one of the four attributes defined in $A'$ or 'None'.

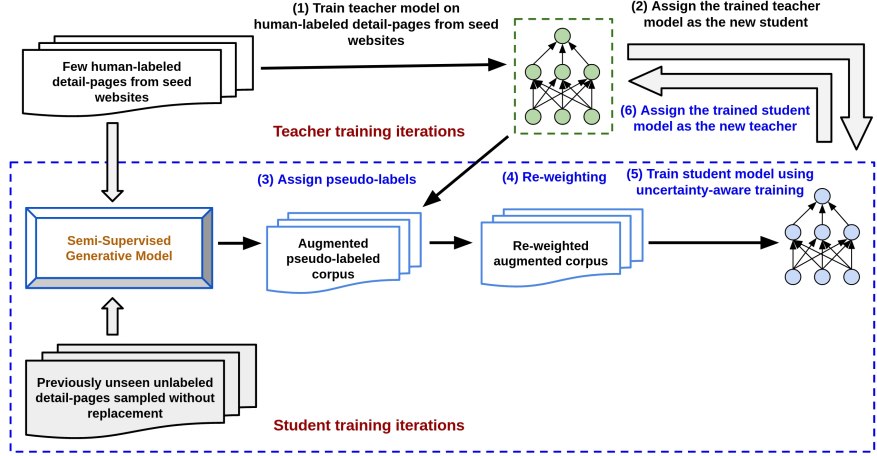

Figure 2: At each iteration, LEAST trains the teacher model on a few human-labeled detail pages from seed websites. The teacher model is then assigned as the new student which trains on an augmented corpus containing both human-labeled and pseudo-labeled samples. The student model is assigned as the new teacher at next iteration.

**Training corpus for DOM node classification.** We represent the training samples derived from a human-labeled detail page as follows: $(X^l = \{x_m^l\}, Y^l = \{y_{m,n}^l\}, P^l = \{p_m^l\}, W^l = \{w_m^l\})_{m=1, n=1}^{N^l, |A'|+1}$, where $x_m^l$ denotes a DOM node, $y_{m,n}^l$ denotes one of the attributes (plus 'None') in the vertical-specific attribute-set $A'$, $p_m^l$ & $w_m^l$ denote the page and the website where the DOM node appears, and $N^l$ represents the total number of DOM nodes appearing on that page. In other words, each training sample in this extraction task corresponds to a DOM node. Similarly, we represent the DOM nodes on an unlabeled detail page as follows: $(X^u = \{x_m^u\}, P^u = \{p_m^u\}, W^u = \{w_m^u\})_{m=1}^{N^u}$, $N^u$ represents the total number of DOM nodes appearing on that page. In our experimental settings, $\Sigma N_u \gg \Sigma N_l$.

**Self-training for DOM node classification.** Recent works [15, 17] have shown that self-training methods can achieve state-of-the-art performance for image/text classification tasks. A traditional self-training framework consists of a teacher model $f(.; \theta_{tea})$ and a student model $f(.; \theta_{stu})$. We initialize the teacher model by training it on some human-labeled samples. The trained teacher model infers pseudo-labels for some task-specific unlabeled data, which is then used to train the student model. Once trained, we assign the student model as the new teacher and repeat this process iteratively until both model converge. For a DOM node classification task, the teacher model $f(.; \theta_{tea}^{(t)})$ assigns each unlabeled DOM node $x_1^u, x_2^u, ..$ in the target vertical a softmax or 'hard' pseudo-label $y_{1,n}^u, y_{2,n}^u, ..$ from the predefined attribute-set $A'$ (plus 'None') after each self-training iteration $t$, $n \in [1, |A'| + 1]$. The student model trains on a corpus that contains both human-labeled & pseudo-labeled samples and updates its parameters as follows.

$$\theta_{stu}^{(t)} = argmin_\theta \frac{1}{N} \Sigma_{m=1}^{N} \mathcal{L}(\hat{y}_{m,n}^{(t)}, f(x_m, \theta_{stu}^{(t-1)})) \quad (1)$$

In Eq. 1, $\mathcal{L}(,)$ can be modeled using cross-entropy-based loss, $x_m$ is a DOM-node, $N$ denotes the size of the corpus, and $\hat{y}_{m,n}^{(t)}$ denotes the label assigned to $x_m$ after $t$-th self-training iteration. We formalize the IE task we study in this paper next.

## 2.1 Problem definition

Given a vertical $v$ (e.g. movie) and some semi-structured websites $W_v = \bigcup_i w_i$ in $v$, where only a few pages from some seed websites $W_v^s$ are labeled by human annotators, and a significantly larger number of detail pages from the same vertical are unlabeled, our objective is to learn a model $f(:, \theta)$ that extracts a predefined attribute-set $A'$ from not only the seed websites but also from previously unseen target websites $W_v^t$ from the same vertical, where $W_v^s \cap W_v^t = \phi$.

## 3 UNCERTAINTY-AWARE SELF-TRAINING

If the number of human-labeled pages is small, initializing the base teacher model by training it on human-labeled samples from scratch would result in a noisy teacher. Using this teacher model to pseudo-annotate unlabeled pages would produce noisy labels. Training the student model on these noisy samples would propagate errors and gradually worsen the model's end-to-end performance [25, 35]. We address these challenges by following a two-phased approach. *First*, we introduce a supplementary supervision source in the form of a *semi-supervised generative model* (Section 3.1). We infer high-quality pseudo-labels for an unlabeled detail page by combining two supervision sources – (i) the generative model, and (ii) the teacher network. *Second*, we mitigate error propagation due to training on noisy samples by re-weighting each sample based on an estimated label accuracy (Section 3.2) and incorporating these weights as a measure of uncertainty during training (Section 3.3).

## 3.1 Semi-Supervised Generative Modeling

We develop a generative model as a supplementary supervision source for pseudo-annotating unlabeled detail pages. Following prior work [3], our model builds on the assumption that a website publishes detail pages by applying a site-specific HTML template to a noisy, partial view of a vertical-specific abstract relation $\mathbb{H}$. In this construct, attribute extraction boils down to simply inverting this generative process to discover the website-specific relation. Assuming partial overlap between websites within a vertical, these relations can be used as a distant supervision source for pseudo-labeling unseen websites in the target vertical.

**Formalization.** Assuming each website $w_i$ is the result of a generative process applied on a vertical-specific abstract relation $\mathbb{H}$, we obtain the detail pages of $w_i$ as follows.

$$w_i = R_i(e_i(\pi_i(\sigma_i(\mathbb{H})))) \quad (2)$$

In Eq. 2, $\sigma_i$ represents the selection operator – it returns a relation containing a subset of the tuples in $\mathbb{H}$; $\pi_i$ represents the projection operator – it returns a subset of attributes; $e_i$ represents website-specific noise – it returns a relation with extraneous attributes and/or replaces attribute values with null/wrong values[1]; $R_i$ represents a website-specific rendering function – it encodes each tuple of the relation $\mathbb{H}_{w_i} = e_i(\pi_i(\sigma_i(\mathbb{H})))$ as a detail page in $w_i$. We present an overview of this generative process in Appendix A.

**Modeling objective.** Given a few human-labeled pages from some seed websites and a set of attributes $A'$ as input, the generative model $\gamma$ outputs a website-specific relation $\mathbb{H}_{w_i}$ for each seed website. We use these relations as a distant supervision source for pseudo-labeling unlabeled pages from the target vertical. We formulate the task of inferring $\mathbb{H}_{w_i}$ as learning a set of website-specific *inverse rendering rules*. An inverse rendering rule is a one-to-one mapping between an attribute $a \in A'$ and a set of DOM nodes in $w_i$. We follow a two-pronged approach to learn inverse rendering rules for each seed website. They are as follows.

**A. Inferring rules from page-level consistency.** Following prior works [26, 33], we encode page-level consistencies in the form of heuristics-based weakly supervised functions. Each function take as input: (i) an attribute ($a \in A'$), (ii) an unlabeled detail page ($p$) & (iii) a list of DOM nodes (*node_list*) labeled as true instances of $a \in A'$ by human annotators, and outputs a list of DOM nodes from $p$. Each function is *sound* with respect to the input attribute $a$, which means if applied to a human-labeled page of a seed website, it will output a subset of DOM nodes that were human-labeled as instances of $a$. These functions can range from regular expressions encoding textual or XPath-based patterns, open-source libraries [21], to functions that combine multiple modalities. The number and type of functions depend on the website and vertical. We provide an example function used in our experiments below.

```
# Fuzzy string matching
    def fuzzy_string_matcher(node_list,a,p){
        pseudo_labeled_list = []
        for node in p.dom_tree():
            if node.hasText():
                for n in node_list:
                    if fuzzy_match(n.text(),node.text()):
                        pseudo_labeled_list.append(node)
        return pseudo_labeled_list
    }
```

**Example 3.1.1.** For each human-labeled node in *node_list*, this function performs a fuzzy match against all DOM nodes in the unlabeled page $p$ that has a text attribute. It considers a DOM node in $p$ a true instance of $a$ if its node-text 'matches' against at least one DOM node in *node_list*. A pair of DOM nodes 'match' if: (a) they both have the same number of words, and (b) the minimum edit-distance between all word-pairs is less than or equal to 3.

---

[1] we make the simplistic assumption that noise is consistent at page-level

**B. Inferring rules from overlapping attribute values.** *Web-Extraction and Integration of Redundant data* or WEIR [3] infers the website-specific relation $\mathbb{H}_{w_i}$ for each seed website by aligning partially overlapping detail pages. It identifies overlapping entities appearing in different websites to align detail pages, and infer a set of rules that map clusters of semantically similar entities to unique identifiers using an unsupervised, hierarchical algorithm. We bootstrap these mappings using a few human-labeled pages to obtain inverse rendering rules for a seed website. We point the readers to the original work by Bronzi et al. [3] for more background on WEIR. We use the official implementation (https://github.com/crescenz/weir) of WEIR in our experiments.

**Pseudo-labeling.** We combine the inverse rendering rules learned from this two-pronged approach to obtain a website-specific relation for each seed website. We use these relations as a distance supervision source to pseudo-annotate detail pages from the target vertical. Specifically, if a DOM node with text $t$ is labeled as an instance of attribute $a \in A'$ in a relation $\mathbb{H}_{w_i}, \exists w_i$, we label all DOM nodes with text $t$ in an unlabeled detail page as an instance of attribute $a$. If a node-text $t$ is assigned different pseudo-labels in different website-specific relations $\mathbb{H}_{w_i}$ and $\mathbb{H}_{w_j}, i \neq j$, we identify the pseudo-label for $t$ by majority voting. We isolate the contribution of inverse rendering rules learned by WEIR and the weakly supervised functions in an ablation study in Section 4.

### 3.2 Node-level Adaptive Re-Weighting

Training a model on noisy labels results in gradual drifts [25] in the model's quality, which degrades the model's end-to-end performance. To mitigate this, we re-weight each DOM node in the training corpus based on its estimated label accuracy. We assign smaller weights to samples with lower accuracy and larger weights to samples with higher accuracy. Consider the pseudo-labels $\{\hat{y}_{m,n}^{(t)}\}_{n=1}^{N_u}$ inferred for the $m$-th unlabeled detail page after $t$-th self-training iteration, $N_u$ represents the total number of DOM nodes on that page. For a corpus with $L$ detail pages, each with $N^u$ DOM nodes, we update the parameters of the student model in a traditional self-training framework using an inner step-size $\alpha$ as follows.

$$\theta_{stu}^{(t)} = \theta_{stu}^{(t-1)} - \alpha \nabla (\frac{1}{L} \frac{1}{N^u} \Sigma_{m=1}^{L} \Sigma_{n=1}^{N^u} \mathcal{L}(\hat{y}_{m,n}^{(t)}, f(x_{m,n}^u, \theta_{stu}^{(t-1)}))) \quad (3)$$

In Eq. 3 $\mathcal{L}(,)$ is a cross-entropy-based loss function and $\nabla(.)$ represents the gradient operator. To account for noisy samples, we leverage the idea of weight perturbation [24] and assign each DOM node a sample weight $c_{m,n}^{(t)} \in (0, 1]$. The key intuition here is to amplify or dampen the loss computed for a training sample by multiplying it with a weight based on its estimated label accuracy. We assign higher weights to samples with higher label accuracy. We assign equal weights to all DOM nodes in a detail page. For a corpus with $L$ detail pages, each with $N^u$ DOM nodes, a weight-perturbed student model is updated as follows.

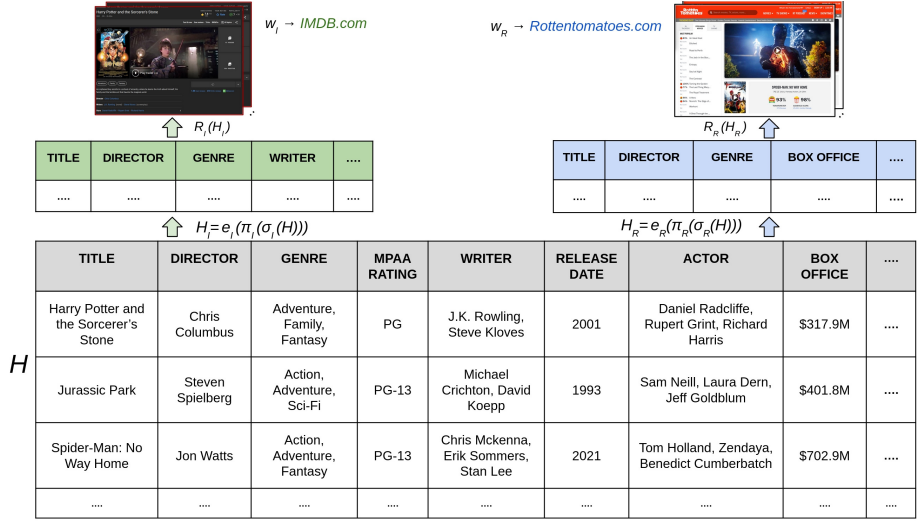

Figure 3: The generative model renders a semi-structured website from a noisy partial view of an abstract relation $\mathbb{H}$ using website-specific HTML templates

$$\theta_{stu}^{(t)} = \theta_{stu}^{(t-1)} - \alpha \nabla \left( \frac{1}{L} \frac{1}{N^u} \Sigma_{m=1}^{L} \Sigma_{n=1}^{N^u} [c_{m,n}^{(t)} \times \mathcal{L}(\hat{y}_{m,n}^{(t)}, f(x_{m,n}^u, \theta_{stu}^{(t-1)})))] \right) \quad (4)$$

**Adaptive weight computation.** We discuss our weight computation strategy in this section. Let, a DOM node $x_i^V$ appears on a detail page $p_i^V$ of website $w_i^V$ in the validation set. The validation set contains human-labeled DOM nodes from the seed websites. At each self-training iteration $t$, we infer a softmax or 'hard' pseudo-label for each DOM node in the validation set. Suppose $\hat{y}_{i,j}^{V^{(t)}}$ represents the softmax pseudo-label assigned to a DOM node $x_i^V$ after the $t$-th self-training iteration, where $j \in [1, |A'|+1]$ and $A'$ represents the attribute-set defined for the extraction task. Suppose $y_{i,j}^l$ denotes the label assigned to this DOM node by a human annotator. We represent the validation set $V^{(t)}$ after the $t$-th self-training iteration as follows: $V^{(t)} = (X^V = \{x_i^V\}, \hat{Y}^{V^{(t)}} = \{\hat{y}_{i,j}^{V^{(t)}}\}, Y^V = \{y_{i,j}^l\}, P^V = \{p_i^V\}, W^V = \{w_i^V\})_{i,j}$.

Let, $(x_m, \hat{y}_{m,n}^{(t)}, p_m, w_m)$ represents a DOM node $x_m$ in the student training corpus appearing on a detail page $p_m$ of website $w_m$. It is assigned a softmax label $\hat{y}_{m,n}^{(t)}, n \in [1, |A'|+1]$ after the $t$-th self-training iteration. To compute its sample weight, we check if $x_m$ appears on a human-labeled detail page of a seed website first. If yes, we assign it a weight of 1.0. Otherwise, we compute its weight based on: (i) the page & the website it appears on, and (ii) its page-overlap against the validation set $V^{(t)}$. If $x_m$ appears on a seed website but the detail page it appears on is not human-labeled, we assign it a weight equal to the average accuracy (normalized within (0,1]) of all 'hard' pseudo-labels inferred for DOM nodes appearing on that specific seed website from the validation set. Otherwise, we assign it a sample weight equal to the average accuracy (normalized within (0,1]) of all softmax pseudo-labels inferred for DOM nodes appearing on the detail page $p_{max}^v \in V^{(t)}$, multiplied by the page-overlap between $p_{max}^v$ and $p_{max}$, where $p_{max}$ represents the detail page from the same website as $x_m$ that has the highest page-overlap against the page $p_{max}^v$ in $V^{(t)}$. We compute page-overlap between a pair of detail pages $p_1$ and $p_2$ as follows.

$$\text{po}_{1,2} = max(\epsilon, \text{JS}(\cup_{i=1}^{|A'|+1} \{s_{i,1}\}, \cup_{j=1}^{|A'|+1} \{s_{j,2}\})) \quad (5)$$

In Eq. 5, JS(.) denotes the Jaccard Similarity, $\{s_{i,j}\}$ represents the set of DOM nodes on page $p_j$ that is assigned the softmax label $i \in [1, |A'| + 1]$, $\epsilon > 0$ is a small number.

### 3.3 Iterative Student-Teacher Training

We start by training the teacher model from scratch on human-labeled samples from the seed websites. Once training terminates, we initialize the student model as a copy of the teacher. We train the student model on a corpus that contains both human-labeled and pseudo-labeled samples from the target vertical. To account for noisy pseudo-labels, we combine two supervision sources – the generative model (Section 3.1) and the teacher model. We assign each sample in the student training corpus a weight $c \in (0, 1]$ based on its estimated label accuracy, noisy samples are assigned lower weights. We train the student model on this re-weighted corpus using a noise-robust loss function (Eq. 7). Once training terminates, we initiate the next self-training iteration by copying the trained student model as the new teacher. We repeat this process until both model converge or a termination criteria is satisfied. We evaluate the performance of the final student model. We provide an overview of our self-training algorithm in Algorithm 1 and present the values for each hyperparameter in the algorithm in Appendix A.

**Pseudo-labeled corpus construction.** At each iteration $t$ of our self-training algorithm, we train the student model $f(:, \theta_{stu}^{(t)})$ on an augmented corpus. We construct this corpus by randomly sampling a maximum of $L$ unlabeled DOM nodes without replacement, assigning each DOM node a softmax pseudo-label, and augmenting

**Algorithm 1** LEAST Algorithm

1: **Input:** Human-labeled DOM nodes ($X^l = \{x_m^l\}, Y^l = \{y_{m,n}^l\}, P^l = \{p_m^l\}, W^l = \{w_m^l\}$); Unlabeled DOM nodes ($X^u = \{x_m^u\}, P^u = \{p_m^u\}, W^u = \{w_m^u\}$); Self-training iterations $T$; Maximum unlabeled sample size $L$; Generative model $\gamma$
2: **Output:** Final student model $f(:, \theta_{stu}^{(T)})$
3: Initialize augmented corpus $A^{(0)}$ and validation set $V^{(0)}$ as mutually exclusive and exhaustive subsets of $(X^l, Y^l, P^l, W^l)$
4: **for** $i = 1$ to $T$ **do**
5:    Train teacher model $f(:, \theta_{tea}^{(t)})$ on $(X^l, Y^l, P^l, W^l)$ ▷ Eq. 3
6:    Sample $L$ DOM nodes from $(X^u, P^u, W^u)$ w.o. replacement
7:    **for** each $x_m^u \in X^u$ **do**
8:      Infer hard pseudo-label $\hat{y}_{m,n}^{(t)}$ from $\gamma$ and $f(:, \theta_{tea}^{(t)})$
9:      Update $A^{(t)} = A^{(t-1)} \cup (x_m^u, y_{m,n}^t, p_m^u, w_m^u)$
10:    **end for**
11:    Initialize validation set $V^{(t)} = V^{(0)}$
12:    **for** each $x \in V^{(t)}$ **do**
13:      Infer hard pseudo-label $\hat{y}^{(t)}$ of $x$
14:      Update $V^{(t)}$
15:    **end for**
16:    **for** each $x \in A^{(t)}$ **do**
17:      Compute weight $c$ of $x$ from $V^{(t)}$    ▷ Section 3.2
18:    **end for**
19:    Initialize $f(:, \theta_{stu}^{(t)}) = f(:, \theta_{tea}^{(t)})$
20:    Update $f(:, \theta_{stu}^{(t)})$ by fine-tuning on $A^{(t)}$    ▷ Eqs. 4 and 7
21:    Update the teacher model $f(:, \theta_{tea}^{(t+1)}) = f(:, \theta_{stu}^{(t)})$
22: **end for**

it to the corpus used to train the student model at the previous iteration $f(:, \theta_{stu}^{(t-1)})$. We initialize the student corpus with the same human-labeled samples used to train the teacher model. The number of pseudo-labeled samples in the student training corpus at each iteration is significantly larger than the number of human-labeled samples. We assign a pseudo-label to each unlabeled DOM node by combining two supervision sources– (i) the teacher model $f(:, \theta_{tea}^{(t)})$, and (ii) the semi-supervised generative model $\gamma$. We fuse these two supervision sources using a stochastic operation. At each iteration $t$, we assign a DOM node $x_m$ in the student training corpus a 'hard' pseudo-label $\hat{y}_{\gamma}^{(t)} \in [1, |A'| + 1]$ inferred by the generative model with probability $\beta^{(t)}$, otherwise we assign it the softmax pseudo-label $\hat{y}_{tea}^{(t)}$ inferred by the teacher model $f(:, \theta_{tea}^{(t)})$. In other words, we prioritize the pseudo-labels inferred by the generative model during the first few iterations of the self-training algorithm and the teacher model during the later iterations. We update the hyperparameter $\beta^{(t)}$ at each iteration as follows; $k_{\beta_1}, k_{\beta_2} \geq 0$ are constants.

$$\beta^{(t)} = \beta^{(t-1)} - k_{\beta_1} \times e^{-k_{\beta_2} t} \quad (6)$$

**Noise-robust loss.** In order to incorporate the estimated label accuracy of a training sample as an uncertainty measure, we utilize a noise-robust loss function for training the student model. It computes the loss for a training sample $x$ with a softmax label $y'$ and a pseudo-label $\hat{y}^{(t)}$ as follows.

$$\mathcal{L}_{ua}(\hat{y}^{(t)}, y') = e^{(1-k^{(t)}c^{(t)})} \times \mathcal{L}(\hat{y}^{(t)}, y') + e^{c^{(t)}} \times U(0, 1) \quad (7)$$

In Eq. 7, $\mathcal{L}(,)$ represents cross-entropy-based loss, $c^{(t)}$ represents the sample weight, and $k^{(t)} \geq 1$ represents a penalty term that accounts for distribution shift due to training samples that represents topic entities that do not appear in the validation set, e.g. long-tailed entities. We amplify the loss term computed from such samples to encourage the model to learn from them. The second term is a regularization factor that prevents the model from overfitting. $U(0, 1)$ denotes a uniform random number within $[0, 1]$. We update $k^{(t)}$ after each iteration $t$ as follows; $k_{c_1}, k_{c_2} \geq 0$ are constants.

$$k^{(t)} = k^{(t-1)} - k_{c_1} \times e^{-k_{c_2} t} \quad (8)$$

We update the teacher model using cross-entropy based loss on human-labeled samples. We update the student model using the noise-robust loss with weight perturbation on an augmented corpus. On a corpus with $L$ detail pages each with $N^u$ DOM nodes, we update the student model as follows.

$$\theta_{stu}^{(t)} = \theta_{tea}^{(t)} - \alpha \nabla (\frac{1}{L} \frac{1}{N^u} \Sigma_{m=1}^{L} \Sigma_{n=1}^{N^u} [c_{m,n}^{(t)} \times \mathcal{L}_{ua}(\hat{y}_{m,n}^{(t)}, y')]) \quad (9)$$

In Eq. 9, $y'$ and $\hat{y}_{m,n}^{(t)}$ denote the softmax label and the 'hard' pseudo-label assigned to the $n^{th}$ DOM node of the $m^{th}$ detail page in the training corpus after $t^{th}$ self-training iteration. $\theta_{tea}^{(t)}$ denotes the teacher model used for initializing the student model. We isolate the contribution of noise-robust loss on end-to-end performance in an ablation study in Section 4.

## 4 EXPERIMENTS

We seek to answer three key questions in our experiments. *First*, how well does a LEAST-trained model perform on the seed websites after being trained on only a few detail pages? *Second*, how well does a LEAST-trained model generalize to unseen websites in the same vertical? *Third*, what are the individual contributions of some of the key components of our self-training algorithm on end-to-end performance? The first two questions answer if a LEAST-trained model works well on a real-world scenario we care most about, where only a few human-labeled pages of some seed websites are human-labeled and we want the model to perform well not only on previously unseen pages of the seed websites but also on unseen websites from the same vertical.

To answer the first question, we train a state-of-the-art neural extraction model [36] using LEAST on 9 human-labeled pages[2] from 2 seed websites & thousands of pseudo-labeled pages from 7 websites in the same vertical of the SWDE dataset [13]. We evaluate its performance on 100 held-out detail pages from the same seed websites. We repeat our experiments on 4 separate verticals. Results show that a LEAST-trained model outperforms previous state-of-the-art model on all verticals. To answer the second question, we compare the same model against a number of strong baselines on 3 unseen websites from the same vertical. We observe that a LEAST-trained model generalizes better to unseen websites on all verticals with significantly less number of human-labeled pages required

---

[2]we selected a random number from [3,10], where 3 is the minimum and 10 is the maximum number of human-labeled pages available from a seed website

for training. To answer the third question, we perform an ablation study that isolates and compares the contribution of some of the key components of our algorithm.

### 4.1 Experiment Design

**Dataset.** We use the publicly available Structured Web Data Extraction (SWDE) dataset [13] for evaluating all competing models. We report results on four verticals– *nba-player, auto, movie,* and *university*. Two of them i.e., *nba-player* and *auto* are considered *dense*, which means that the websites in these verticals have higher data overlap, whereas the other two verticals are comparatively *sparse*. Each vertical has 10 websites, thousands of detail pages, and 4 attributes. We present a summary overview of each vertical in Table 1.

**Experiment settings.** We evaluate an extraction model's performance in two different settings: (i) zero-shot setting and (ii) in-domain setting. In both cases, we randomly select 9 human-labeled pages from 2 seed websites to train all models. The validation set contains 10 human-labeled pages from each seed website. For the in-domain setting, we test a model's performance on a corpus that contains 100 held-out detail pages from each seed website. For the zero-shot setting, we test the model's performance on a corpus containing all detail pages from 3 randomly selected, previously unseen, unlabeled target websites from the same vertical. Needless to say that the target websites are different from the seed websites. For both settings, we pseudo-annotate unlabeled pages from 7 websites from the same vertical (barring the target websites in zero-shot setting and the detail pages comprising the test corpus for in-domain setting) to LEAST-train an extraction model. This materializes to 3,092 unlabeled pages from *nba-player*, 13,249 unlabeled pages from *auto*, 13,982 unlabeled pages from *movie*, and 11,660 unlabeled pages from the *university* vertical. We use the same evaluation metrics as the original authors of the SWDE dataset [13] and compute average page-level macro-F1 scores. Following prior work [18, 36], we only consider the top-1 prediction for each attribute during evaluation.

**Table 1: Summary statistics of the experimental dataset**

| Vertical | Websites | Pages | Attributes to be extracted |
|---|---|---|---|
| *nba-player* | 10 | 4,405 | name, team, height, weight |
| *auto* | 10 | 17,923 | model, price, engine, fuel_economy |
| *movie* | 10 | 20,000 | title, director, genre, mpaa_rating |
| *university* | 10 | 16,705 | name, phone, website, type |

### 4.2 Baseline Methods

**M1. Web-extraction using overlapping data (`WEIR`).** WEIR [3] follows an unsupervised algorithm to align detail pages from partially overlapping websites in the same vertical. It infers a list of wrappers that map a group of similar entities to a unique identifier using a hierarchical algorithm. We bootstrap these mappings using the human-labeled detail pages from the seed websites. This provides us with text-spans from the target websites corresponding to each attribute. We use the official implementation in our experiments.

**M2. Visual rendering-based model (`Render-Full`).** Hao et al. [13] utilize visual features to encode the distance between different visual elements in a fully-rendered detail page in a web-browser. Although visual features are strong candidates to encode the relationship between neighboring DOM nodes, they require a time-consuming rendering process as well as additional memory to store images, CSS, and JavaScripts that can easily get out-of-date. We implement `Render-Full` that employs a heuristics-based algorithm leveraging visual features and report strong end-to-end performance than other variants such as `Render-PL` and `Render-IP`.

**M3. Simplified DOM trees (`SimpDOM`).** SimpDOM [36] encodes discrete multimodal features (e.g. leaf node type, xpath, position, text) for each DOM node in a *simplified* DOM tree using a bi-directional LSTM network, followed by a MLP-based softmax classification layer. One of the key contributions of this work is a heuristics-based DOM tree simplification step that prunes redundant DOM nodes in the original DOM tree and encodes contextual features of a node by identifying its most influential neighbors. To encode node-text, they aggregate character-level embeddings of each token using a convolutional network and append it to a GLoVe-based word embedding vector. They train their model using cross-entropy-based loss on the human-labeled pages of seed websites and report state-of-the-art zero-shot performance. We present an illustrated overview of `SimpDOM` in Appendix A. We follow the official implementation of `SimpDOM` in our experiments.

**M4. Relational graph neural network (`GNN`).** We implement a relational graph neural network model consisting of three modules– text encoder, graph propagation, and classification. We represent each DOM node using three textual fields: HTML tag, attributes, and node-text and its relative position among its siblings. We initialize each text field as random word embedding vectors. We sum them with the node's relative position to derive the initial representation vector, which is then propagated through a LSTM layer with multi-headed attention [8]. In order to propagate information from children to parent, as well as between sibling nodes, we introduce edges between each parent-child and sibling pair in the DOM tree. We aggregate information for each DOM node in this modified DOM tree from neighbors within 6 hops using a relational graph convolutional layer [27]. The node representations are then propagated through a LSTM layer, followed by a linear layer and a softmax classification step. We train the model using a cross-entropy-based loss function. We provide an illustrated overview of GNN in Appendix A.

**Implementation details.** We LEAST-train GNN & SimpDOM and compare their performance on zero-shot setting against all baselines. We use the open-source LXML library (https://lxml.de/) to identify the DOM tree structure of a detail page. For training a `SimpDOM`-based model, we follow the official implementation (https://github.com/google-research/google-research/tree/master/simpdom). We use a word embedding and character embedding vectors of size 100, whereas xpath, leaf-type and position embedding vectors are of dimensions 30, 30, & 20 respectively. We use pretrained GLoVe embedding for initializing the word embedding vectors and randomly initialize the rest of the embedding vectors. We use a convolutional network with 50 filters & kernel size of $3 \times 3$, and a bi-directional LSTM network with hidden layer size of 100. We train this model for 15 epochs with a batch size of 32. For fair evaluation, we use the author recommended hyperparameters during training. For training a GNN-based model, we set the maximum length of HTML tag, attributes and node-text to 10, 64, & 64 respectively. We randomly initialize the word embedding vectors and use a relational graph convolution network with 100 filters & kernel size of 3 × 4 × 5. We train the model for 25 epochs with a batch size of 8 and learning rate of 0.001. To prevent overfitting, we apply dropout following the linear layer with a dropout rate of 0.3. We implement this model in PyTorch.

## 4.3 Results and Discussion

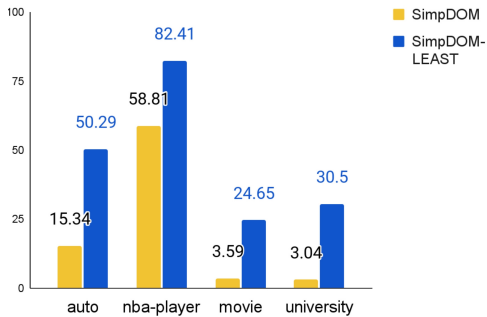

Figure 4: Average F1 score of `SimpDOM` and `SimpDOM-LEAST` on four verticals of the SWDE dataset

**A. Zero-shot setting.** We report page-level average precision, recall and macro-F1 score of all competing models on 3 unseen websites from the same vertical, averaged over 4 verticals in Table 2. Following the original authors' recommendations, we train models M1 to M4 on human-labeled pages from the seed websites. We train all models on the same number of human-labeled pages. Only LEAST-trained models utilize pseudo-labeled samples during training. We observe that `GNN-LEAST` and `SimpDOM-LEAST` perform the best in zero-shot setting. `SimpDOM-LEAST` boosts its average F1-score compared to `SimpDOM` by 27.53%. In case of `GNN-LEAST`, we observe a 32.55% improvement in average F1 score. This shows that LEAST-trained models generalize better on unseen websites. We observe that `SimpDOM-LEAST` performs better than `GNN-LEAST` on all verticals. We hypothesize that this is due to the DOM tree simplification step undertaken by `SimpDOM-LEAST` as part of its data preprocessing subroutine. We also observe (see Fig. 4) that improvement in performance for a LEAST-trained model is higher for dense verticals (e.g. *nba-player*, *auto*) than sparse verticals (e.g. *university*, *movie*). For example, `SimpDOM-LEAST` improves the average F1 score of `SimpDOM` by 21.01% for *movie*, compared to 23.60% for the *nba-player* vertical.

Table 2: Average zero-shot performance on four verticals

| Index | Model | End-to-end performance | | |
|---|---|---|---|---|
| | | *Precision (%)* | *Recall (%)* | *F1 (%)* |
| M1 | WEIR | 7.18 | 2.08 | 3.22 |
| M2 | Render-Full | 32.15 | 11.10 | 16.50 |
| M3 | SimpDOM | 37.22 | 13.85 | 20.19 |
| M4 | GNN | 9.09 | 3.73 | 5.29 |
| M5 | GNN-LEAST | 58.07 | 28.06 | 37.84 |
| M6 | SimpDOM-LEAST | **60.48** | **38.38** | **46.96** |

**B. In-domain setting.** A generalizable web extraction model should perform well on unseen detail pages from the seed websites. We investigate a LEAST-trained model's performance on the seed websites in this section. We compare the performance of `GNN-LEAST` and `SimpDOM-LEAST` against `GNN` and `SimpDOM` on a test corpus that contains 100 held-out detail pages from each of the 2 seed websites. We report page-level average precision, recall and F1-score averaged over 4 verticals in Table 3. We observe best end-to-end performance using `GNN-LEAST`, closely followed by `SimpDOM-LEAST`. Results show that LEAST-training boosts a model's extraction performance on unseen detail pages of the seed websites. We observe an improvement of 16.48% and 10.89% in average F1 score over `GNN` and `SimpDOM` using their LEAST-trained counterparts. We present more results and analysis in Appendix A.

Table 3: Average in-domain performance on four verticals

| Index | Model | End-to-end performance | | |
|---|---|---|---|---|
| | | *Precision (%)* | *Recall (%)* | *F1 (%)* |
| M3 | SimpDOM | 84.54 | 70.90 | 77.13 |
| M4 | GNN | 53.79 | 87.14 | 66.52 |
| M5 | GNN-LEAST | 87.83 | **98.82** | **93.0** |
| M6 | SimpDOM-LEAST | **98.16** | 79.78 | 88.02 |

**C. Label-efficiency.** We investigate the label-efficiency of a LEAST-trained model by reporting the average F1 score of `SimpDOM-LEAST` in zero-shot setting with varying number of seed websites. We train the model on 9 human-labeled detail pages from $n = \{2, 3, 4, 5\}$ seed websites and thousands of pseudo-labeled pages from $7 - n$ websites from the same vertical. We report the average F1 score on 3 unseen websites from the same vertical in Fig. 5. We observe a positive correlation between end-to-end performance with the number of seed websites. We observe higher improvements in F1 score for *nba-player* and *auto* with increasing number of seed websites. We also report the labeling effort saved by LEAST in Table 4 by comparing the number of human-labeled pages needed from 2 seed websites to train a `SimpDOM` model to obtain comparable performance as shown in Fig. 4. We initialize the training corpus with the same number of human-labeled pages and keep increasing the number of human-labeled pages from each seed website by 25 until `SimpDOM` achieves comparable F1 score.

Table 4: Number of human-labeled pages needed to train `SimpDOM` for comparable performance

| Vertical | SimpDOM-LEAST | SimpDOM | Saved |
|---|---|---|---|
| *nba-player* | 18 | 368 | 350 |
| *auto* | 18 | 668 | 650 |
| *movie* | 18 | 818 | 800 |
| *university* | 18 | 218 | 200 |

**D. Ablation study.** We isolate the contribution of some of the key components of the LEAST algorithm and compare their performance against other viable alternatives in Table 5. For each scenario, we report the corresponding average F1 score in zero-shot setting. In S1, we only use the teacher model for pseudo-labeled corpus construction. We observe that not using the generative model as a supplementary supervision source worsens average F1 score by 21.42%. In S2, we only use the inverse rendering functions inferred by WEIR within our generative model for pseudo-labeling. This decreases average F1 score by more than 18 points. In S5 and S6, we train SimpDOM using two state-of-the-art self-training algorithms– mean-teacher [30] and UST [23]. We outperform both by 19 average F1 points. This establishes that simply extending contemporary self-training algorithms do not work well for transferable web extraction if the number of human-labeled pages is small. In S4, we replace the noise-robust loss with a cross-entropy-based loss function for student training in the LEAST-training workflow. We observe a 1.85% drop in average F1 score. Finally, in S3, we assign all training samples equal weight. We observe a 1.51% drop in average F1 score. This shows that re-weighting helps a model generalize better on unseen websites when it is trained on pseudo-labeled samples.

Table 5: Results of the ablation study

| Index | Description | Average F1 (%) |
|---|---|---|
| S0 | SimpDOM-LEAST | 46.96 |
| S1 | SimpDOM-LEAST w.o. generative modeling | 25.54 (↓ 21.42) |
| S2 | SimpDOM-LEAST w. WEIR | 29.95 (↓ 18.01) |
| S3 | SimpDOM-LEAST w.o. adaptive re-weighting | 45.67 (↓ 1.51) |
| S4 | SimpDOM-LEAST w.o. noise-robust loss | 45.11 (↓ 1.85) |
| S5 | SimpDOM w. mean teacher | 26.79 (↓ 20.17) |
| S6 | SimpDOM w. UST | 27.11 (↓ 19.85) |

## 5 RELATED WORKS

Our work builds on recent works from IE and NLP community.

**Attribute extraction.** Attribute extraction from web documents is an extensively studied problem in supervised settings [1, 29]. Early works [12] usually require a large number of human-labeled pages to learn wrappers to identify attributes from each website. Although they yield high accuracy, wrapper induction methods do not generalize well on unseen websites. To address this, Zhai et al. [34] proposed an active learning based approach to find the most important training samples in a target website for labeling and then adapt the existing wrappers to each target website in a cost-effective way. Their approach, however, required significant human-effort in building specialized annotation tools and label new samples from a target website. Lockard et al. [19] used an external knowledge base as a distant supervision source to label text-fields in a target website. A sufficiently large and comprehensive knowledge base, however, may not always be available for emerging domains. To address this, we pseudo-annotate unlabeled pages from the same vertical to construct our training corpus.

**Transferable web extraction.** Hao et al. [13] used visual distance-based features computed from a fully rendered web page to achieve promising results on unseen websites without requiring any new human annotations. Lockard et al. [20] also utilized visual rendering-based features to encode node-level contextual features using a graph neural network. Computing visual rendering-based features, however, is a time-consuming process that require additional memory to store images, CSS, and JavaScripts which can easily get out-of-date. Lin et al. [18] addressed these by developing a relational neural network that encodes pairwise relationship between neighboring DOM-nodes that generalize well on unseen websites without requiring any visual rendering. We implemented a relational graph neural network and compared its end-to-end performance in zero-shot setting after training it only on human-labeled pages similar to [18] and after LEAST-training in Table 2. We observed significant improvement in performance after LEAST-training. Zhou et al. [36] introduced some heuristics to prune the neighborhood of a DOM-node, and identify the most influential neighbors. They represented each DOM-node using a number of discrete features and classified them as one of the attributes to be extracted. However, when trained on a few human-labeled pages, the transferability of their model degrades significantly as we observe in Table 2.

**Semi-supervised training.** Semi-supervised training have been successfully employed for many instance-level classification tasks [22, 23, 30] in recent years. We observe that a vanilla extension of these techniques do not work well (see Table 5) for web-extraction when the number of human-labeled samples is small. Latent variable modeling methods such as SeqVAT [5] and CVT [6] do not work well in limited labeled data settings [32] as they ignore model uncertainty resulting from very few human-labeled samples. Self-training [28] is one of the earliest semi-supervised approaches to train a model on limited labeled data. Self-training with noisy pseudo-labels is an active area of research [10]. A major line of works [14] focus on correcting noisy labels by learning label corruption matrices. More related to our work, however, are instance re-weighting approaches [15, 23, 32] that down-weight samples with noisy labels. In contrast to these, we do not rely on resource-intensive language models, nor do we rely on prior knowledge about the content of the unlabeled samples.

## 6 CONCLUSION

Conventional web extraction methods such as wrapper induction yield high accuracy but do not generalize well on unseen websites. Prior works on transferable web-extraction trains their models on a large number of human-labeled detail-pages from seed websites. We propose LEAST, a label-efficient self-training framework that trains transferable web-extraction models using only a few human-labeled and a large number of pseudo-labeled pages from the same vertical. LEAST works in tandem with multiple existing web-extraction models. Through exhaustive experiments on a large-scale publicly available dataset, we show that a LEAST-trained model not only generalizes better on previously unseen websites than the previous state-of-the-art, but it also reduces the amount of human-effort required to label training samples to achieve comparable performance.

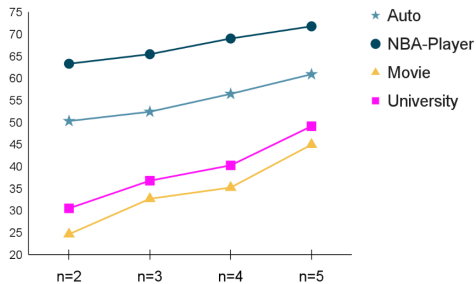

Figure 5: Average F1 score (%) achieved by SimpDOM-LEAST using varying number of seed websites in zero-shot setting

## 7 APPENDIX A

**Table 6: Values of different hyperparameters used for self-training using LEAST framework**

| Symbol | Usage | Value |
| --- | --- | --- |
| $T$ | Maximum no. of self-training iterations | 5 |
| $L$ | Maximum sample size for unlabeled DOM nodes | 100,000 |
| $\beta^{(0)}$ | Pseudo-labeled corpus construction | 0.6 |
| $k_{\beta_1}$ | Update $\beta^{(t)}$ | 0.1 |
| $k_{\beta_2}$ | Update $\beta^{(t)}$ | 1 |
| $k^{(0)}$ | Penalty term in noise-robust loss | 1 |
| $k_{c_1}$ | Update $k^{(t)}$ | 0.1 |
| $k_{c_2}$ | Update $k^{(t)}$ | 1 |
| $\epsilon$ | Page overlap computation | 0.0005 |

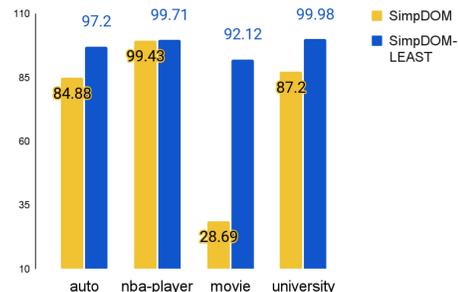

**Figure 6: Average F1 score over four verticals in 5-shot setting**

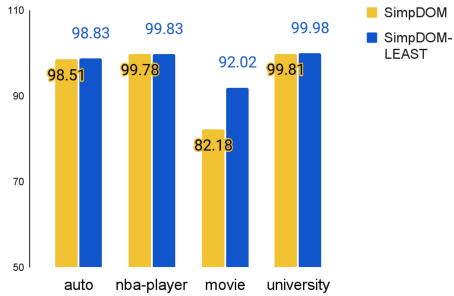

Figure 7: Average F1 score over four verticals in 100-shot setting

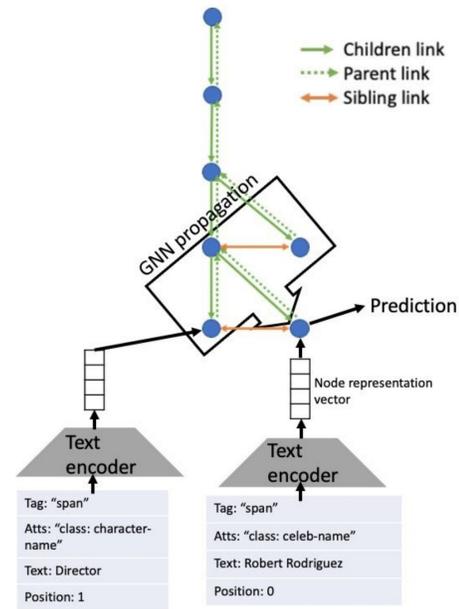

Figure 9: Overview of our Relational Graph Neural Network (GNN)

*Fig. 9 shows an overview of our relational graph neural network. It encodes each DOM node using its HTML tag, attribute node-text. and relative position. Aggregate information is accumulated from neighbors using a relational graph convolution layer with multi-headed attention. Each node is categorized as one of the attributes to be extracted (plus 'None') using a softmax classifier.*

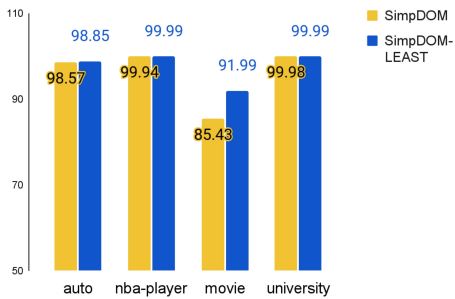

Figure 8: Average F1 score over four verticals in 400-shot setting

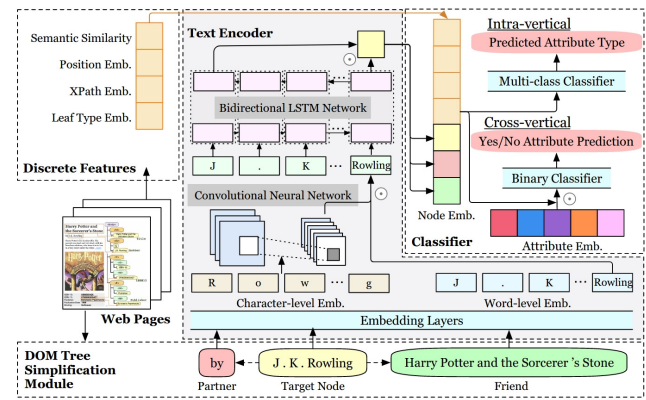

Figure 10: Overview of SimpDOM's model architecture from Zhou et al. [36]

*Fig. 10 shows an overview of the SimpDOM model architecture. It encodes each DOM node's textual features using LSTM and CNN at the word-level and character-level respectively and then concatenates it with a set of discrete features built from the DOM trees including leaf type, XPath, and the relative position of each node. They develop a heuristics-based tree simplification method that prunes redundant*

We compare the performance of SimpDOM-LEAST against SimpDOM on three separate $k$-shot settings, where $k$=5, 100, 400. For each setting, we randomly select $k$ human-labeled pages from 3 target websites along with 9 human-labeled pages from 2 seed websites. We report both models' average page-level F1-score on a test corpus containing 100 held-out detail pages from each target website in Fig. 6, 7, and 8. For LEAST-training, we pseudo-annotate all unlabeled pages from a vertical barring the pages that constitute the test corpus.

*DOM nodes and edges to identify the most influential neighbors of a DOM node in a detail page (see Fig. 1 for an example), referred as 'partner' (by) and 'friends' (Harry Potter and the Sorcerer's Stone) of a node (J. K. Rowling). These features are used to represent each node in a classification task for extracting a predefined set of attributes.*